# Service Quality Improvement of Mobile Users in Vehicular Environment by Mobile Femtocell Network Deployment

Mostafa Zaman Chowdhury[a,£] and Yeong Min Jang[a,*]
[a] Department of Electronics Engineering, Seoul 136-702
Kookmin University, Korea
E-mail: [£]*mzceee@yahoo.com*, [*]*yjang@kookmin.ac.kr*

*Abstract*— The femto-access-point (FAP), a low power small cellular base station provides better signal quality for the indoor users. The mobile users in the vehicular environment suffer for the low quality signal from the outside wireless networks. The deployment of femtocells in the vehicular environment can solve the low level signal-to-noise plus interference problem. In this paper, we propose new application of the femtocell technology. The femtocells are deployed in the vehicular environment. Short distance between the user and the FAP provides better signal quality. The inside FAPs are connected to the core network through the outside macrocellular networks or the satellite networks. One stronger transceiver is installed at outside the vehicle. This transceiver is connected to FAPs using wired connection and to macrocellular or the satellite access networks through wireless link. The capacity and the outage probability are analyzed. The simulation results show that the proposed mobile femtocell deployment significantly enhances the service quality of mobile users in the vehicular environment.

## I. Introduction

The wireless engineering has been searching for low-cost indoor coverage solutions since the beginning of mobile networks. Femtocellular network technology is one of such solutions. The femto-access-points (FAPs) are low-power, small-size home-placed Base Stations (also known as Home NodeB or Home eNodeB) that enhance the service quality for the indoor mobile users [1]-[3]. Some key advantages of femtocellular network technology are the improved coverage, reduced infrastructure and the capital costs, power saving, improved signal-to-noise plus interference (*SNIR*) level at the mobile station (MS), and improved throughput. Femtocells operate in the spectrum licensed for cellular service providers [2], [4], [5]. Thus, it can provide high QoS. Also, no need to use the dual-mode terminal for this technology, whereas WLAN needs dual mode terminal. Thus, the key feature of the femtocell technology is that users require no new equipment (UE).

Currently the mobile users inside a vehicular environment use macrocellular or the satellite networks. However, the users suffer from several difficulties e.g., low *SNIR* level, higher outage probability, and lower throughput due to the poor signal quality inside the vehicle. Femtocells deployment in the vehicular environment, we refer as the mobile femtocell can solve these difficulties. The MS is connected to the indoor FAP instead of outside macrocellular or the satellite networks. A strong transceiver is installed outside the vehicle. The outside transceiver is connected to the inside FAP by a wired network and with outside macrocellular or the satellite network through a wireless link. Thus, the signal does not need to penetrate the wall of the vehicle. Therefore, the MS can receive better quality signal. Repeater (or signal booster) can also solve the poor coverage issues in vehicular environment. However, data speed of the Repeated is very low compared to femtocells.

Several standardization bodies e.g., Femto Forum [5], 3GPP [6], and 3GPP2 and some research groups e.g., AirWalk, ABIresearch, Kineto wireless, picoChip flexible wireless, etc. are working to improve this technology. 3GPP is focusing on the standardization of the femtocell architecture for UMTS (W-CDMA) and 3G Long-Term Evolution (LTE). 3GPP2 is focusing on the standardization of the femtocell architecture for CDMA2000 (EV-DO) and Ultra-Mobile Broadband (UMB). DSL Forum is focusing on the extension of the TR-069 *management* protocol to support femtocells. Femto Forum is focusing on the inclusive organization dedicated to the promotion of femtocells. However, the research on mobile femtocells i.e., the femtocellular network deployment in vehicular environment is not focused by any researcher yet. We propose that the inclusion of research on mobile femtocell will open the new door of femtocell technology. The simulation results show that the proposed scheme is able to improve the service performances significantly of a mobile user in a vehicular environment.

The rest of this paper is organized as follows. Section II shows the system model for the proposed scheme. Service scenarios for the mobile femtocellular network deployment are shown in Section III. Section IV provides the capacity and outage probability analyses. Finally, conclusions are drawn in the last section.

## II. System Model for Mobile Femtocellular Network Deployment

In this section, we have sown the system model to deploy the femtocellular networks inside a vehicle. A FAP is located inside the vehicle e.g., bus/car/train/ship or others. A transceiver is situated outside the vehicle to transmit/receive data to/from the backhaul networks (e.g., macrocellular networks, satellite networks, and etc.). The FAPs are installed inside the vehicles to make wireless connection between the users and the FAP. The FAPs and the transceiver are connected through the wired networks. The overall FAP-to-core network connectivity is shown in Fig. 1. In this system, the FAP works like a relay station (RS). In the proposed model, the existing macrocellular networks or the satellite networks are used for backhauling the femtocell traffic of the mobile users inside the vehicle. The antenna of the outside transceiver is relatively stronger compared to the antenna of the MS. The better quality received signal from FAP provides enhanced quality of services in terms of capacity, signal quality, and outage probability.

---
[*]Corresponding author

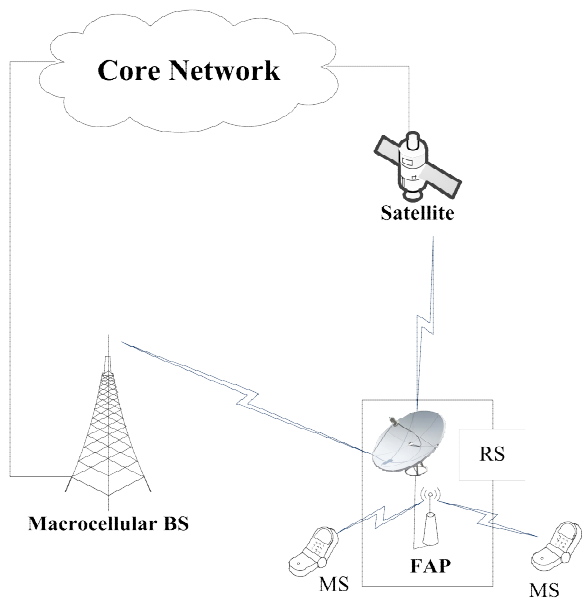

**Fig. 1**. The basic connectivity for the FAP to core network for the mobile femtocells deployment.

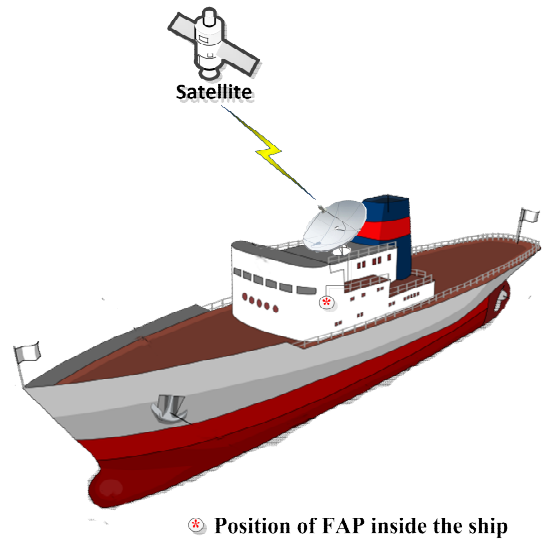

**Fig. 2.** Service scenario for the mobile femtocells deployment in a ship.

## III. Proposed Service Scenarios

The service scenario is important issue before deploying a new technology. The mobile femtocell is the new paradigm of the femtocellular network technology. The signal quality inside a vehicular environment is not good enough whenever the MS directly connected with macrocellular or others wireless networks. Especially if the distance between the vehicle and the access network is far, the user suffers from several difficulties, e.g., increased outage probability, reduced *SNIR*, and reduced throughput. Therefore, mobile femtocells can be deployed in various vehicular environments to enhance the service quality. The approach of network configuration should be different for the different vehicular environments due to the speed of vehicles and the availability of wireless backhaul networks. Some service scenarios of mobile femtocells networks are proposed here.

### 3.1 Mobile Femtocells in a Ship

In a ship environment, the subscribers normally stay inside the ship. To enhance the service quality, femtocells can be deployed in the ship. In this case, the ship is normally out of macrocellular network coverage area. Hence, the network used for backhauling the femtocell traffic must be satellite network. The macrocellular network can be used for backhauling the femtocell traffic only when the ship stays at port where the macrocellular coverage is available. Fig. 2 shows the scenario of mobile femtocells in a ship environment. The outside transceiver receives the satellite signal. The FAPs inside the ship are connected to the outside transceiver through the wired connection. Then the MS receives the better quality signal from the FAP.

### 3.2. Mobile Femtocells in a Slow or Medium Speed Vehicle

In a slow or medium speed vehicle, the deployment of mobile femtocells can improve the service quality. For the slow or medium speed cases, both the macrocellular networks and the satellite networks can be used for backhauling the femtocell traffic. Normally the satellite access network is very expensive compared to the macrocellular networks. The satellite networks should be chosen as a secondary option. Therefore, a femtocell network is connected to the satellite networks only when the macrocellular networks is not available. Fig. 3 shows a scenario of the mobile femtocell network deployment in this scenario. The train of medium speed moves around the country. Hence, there is some possibility that macrocellular network coverage is not available. If the macrocellular network is not available then for that duration, the femtocell network is connected to the satellite networks for the backhauling of femtocells traffic. On the other hand, if the vehicle only moves around the city where the macrocellular coverage is available everywhere, then there is no need of the satellite connectivity. Fig. 4 shows a service scenario for the mobile femtocells deployment in a slow or medium speed vehicle where only the macrocellular network is provided. The bus is only moved around the city. Hence, the macrocellular network coverage is always available to provide the service connectivity.

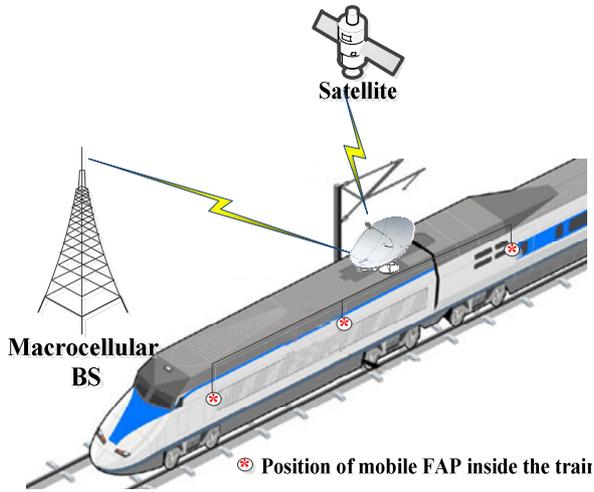

**Fig. 3.** Service scenario for the mobile femtocells deployment in a slow or medium speed vehicle where both the satellite and the macrocellular networks are provided.

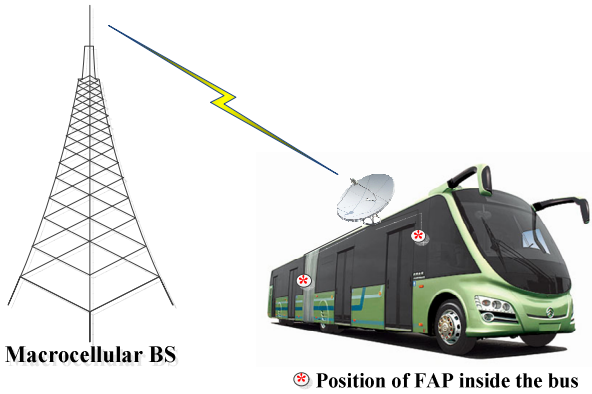

**Fig. 4.** Service scenario for the mobile femtocells deployment in a slow or medium speed vehicle where only the macrocellular network is provided.

*3.3 Mobile Femtocells in a High Speed Vehicle*

In a fast speed vehicle, mobile femtocells can also be deployed to enhance the service quality for the mobile users. However, for this case macrocellular network is not effective for backhauling the femtocell traffic. Because the vehicle is very fast and macrocellular network will make very frequent network handover. Thus, the performance will be degraded due the frequent handovers. Therefore, the satellite network is the best possible solution for the backhauling of traffic from femtocellular networks. The femtocells can be connected to the macrocellular networks only when the vehicle is in stationary or speed of the vehicle is less than a threshold speed. Fig. 5 shows a scenario of the mobile femtocell network deployment in a high-speed train. The train moves very fast. So, during the moving time with high speed, the femtocells are connected to the satellite networks only.

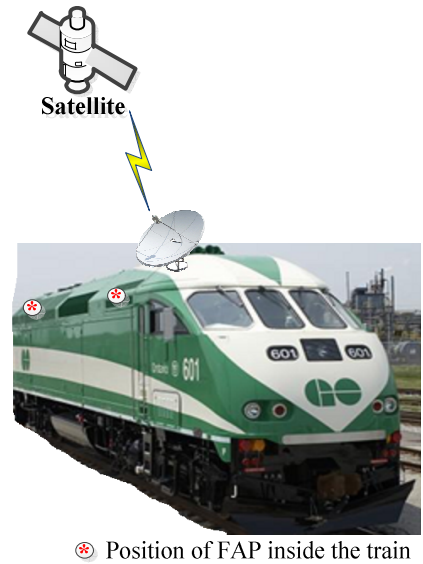

**Fig. 5.** Service scenario for the mobile femtocells deployment in a high-speed vehicle where only satellite network is provided for the FAPs connection.

## IV. Outage Probability and Capacity Analyses

There are various interference mechanisms present in the macrocell/femtocell integrated network architecture; in particular, between macrocells and femtocells, and among femtocells. In addition, the noise is an element of the wireless environment. These interferences and the noise affect the capacity of the wireless link and also the outage probability of a user from the link. In this section, we provide the capacity and the outage probability analyses for the mobile femtocellular network deployment.

The propagation model [7], [8] for a macrocell case can be expressed as:
the user in the outdoor environment,

$$L = 36.55 + 26.16 \log_{10} f_{c,m} - 3.82 \log_{10} h_b - a(h_m) \\ + [44.9 - 6.55 \log_{10} h_b] \log_{10} d + L_{sh} \quad [dB], \quad (1)$$

the user in the vehicular environment,

$$L = 36.55 + 26.16 \log_{10} f_c - 3.82 \log_{10} h_b - a(h_m) \\ + [44.9 - 6.55 \log_{10} h_b] \log_{10} d + L_{sh} + L_{pen} [dB] \quad (2)$$

$$a(h_m) = 1.1[\log_{10} f_{c,m} - 0.7]h_m - (1.56 \log_{10} f_{c,m} - 0.8) \quad (3)$$

where $f_{c,m}$ is the center frequency in $MH_Z$ of the macrocell, $h_b$ is the height if the macrocellular BS in meter, $h_m$ is the height of the MS in meter, $d$ is the distance between the macrocellular BS and the MS in kilometer, $L_{sh}$ is the shadowing standard deviation, and $L_{pen}$ is the penetration loss.

The propagation model for a femtocell case can be expressed as:
the users receiving the signal form the FAP that is located inside the same vehicle,

$$L_{fem} = 20\log_{10} f_{c,f} + N\log_{10} d_1 - 28 \quad [dB], \quad (4)$$

the users receiving the signal form the FAP that is located outside the vehicle

$$L_{fem} = 20\log_{10} f_{c,f} + N\log_{10} d_1 + 4n^2 - 28 \quad [dB] \quad (5)$$

where $f_{c,f}$ is the center frequency in MHz of the femtocell, $n$ is the number of walls between the MS and the FAP, and $d_1$ is the distance between the FAP and the MS in meter.

Assuming that the spectrums of the transmitted signals are spread, we can approximate the interference as AWGN. Then, following the Shannon Capacity Formula,

$$C = \log_2(1 + SNIR) \quad [bps/H_z] \quad (6)$$

The capacity of a wireless channel decreases with decreased signal-to-noise plus interference (*SNIR*) level. The received *SNIR* level of a femtocell user in a macrocell/femtocell integrated network can be expressed as:

$$SNIR = \frac{S_0}{I_f + I_m + N} \quad (7)$$

where $S_o$ is the power of the signal from the associated macrocellular BS or FAP, $I_m$ is the total power of the interference signals from all of the interfering macrocells, $I_f$ is the total power of the received interference signals from all of the interfering femtocells, and $N$ is the total power of the received noise.

The outage probability [4], [9] of a user is calculated as:

$$P_{out} = P_r(SINR < \gamma) \quad (8)$$

where $\gamma$ is a threshold value of *SNIR* below which there is no acceptable reception.

Considering all the interfering neighbor macrocells and femtocells, the outage probability can be expressed as:

$$P_{out} = 1 - e^{-\frac{\gamma}{S_o}(I_f + I_m + N)} \quad (9)$$

Considering the macrocell, to transit signal for a MS inside the vehicle, the signal needs to penetrate the body of a vehicle. Thus, the signal level degraded significantly inside the vehicle. On the other hand, the FAP and the MSs are very close. So, the received signal level is very good. Therefore, the *SNIR* level can be improved using the mobile femtocellular network deployment. The link of the FAP-to-MS is also depended on the link between the macrocellular BS and the outside receiver. However, the received signal at the outside receiver is much stronger than the received signal at MS inside the vehicle from the macrocellular BS. Hence, the improved *SNIR* through the mobile femtocellular network deployment provides better link throughput and the outage probability can reduced significantly.

## V. Simulation Results

In this section, we evaluate the capacity and the outage probability of the proposed mobile femtocellular network deployment. Table 1 summarizes the values of the parameters that we used in our analysis. In the vehicular environment, the interference from the other femtocells was not considered. We consider the macrocellular networks as the backhauling networks. Also, we assume that the femtocells in the vehicles and the overlaid macrocell are deployed through separate frequency bands [2]. We took the average of 100 separate simulation results.

**Table 1:** Summary of the parameter values used in our analysis

| Parameter | Value |
|---|---|
| Distance between the FAP and the MS | 5 [m] |
| Carrier frequency | 1800 [MHz] |
| Transmit signal power by the macrocellular BS | 1.5 [kW] |
| Transmitted signal power by a FAP | 15 [mW] |
| Height of a macrocellular BS | 100 [m] |
| Height of a FAP | 2 [m] |
| Threshold value of *SNIR* for MS ($\gamma$) | 10 |
| Threshold value of *SNIR* for outside transceiver ($\gamma$) | 7 |
| $L_{pen}$ | 20 [dB] |
| $L_{sh}$ | 8 [dB] |

Fig. 6 shows the comparison between the *SNIR* levels. The signal level at MS from the FAP is almost same because the distance between the MS and the FAP is constant. However, the signal level decreases with the increase of the distance between the macrocellular BS and the MS. Hence, the *SNIR* level becomes very poor when distance between the macrocellular BS and the vehicle is more than 500 m. The figure clearly shows the advantage of the mobile femtocellular networks especially when the distance between the macrocellular BS and the MS is comparatively higher.

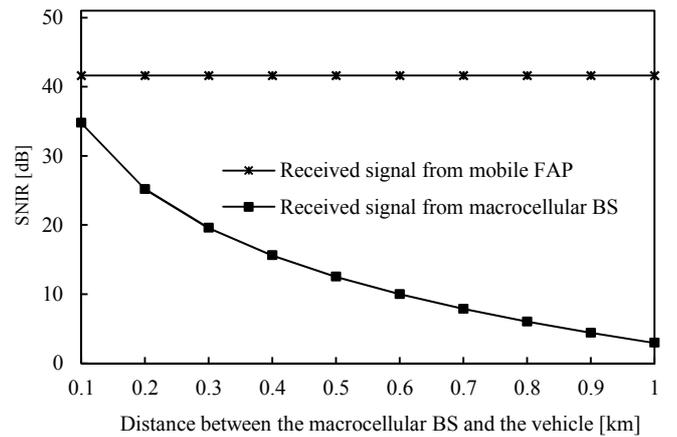

**Fig. 6.** Comparison of *SNIR* levels between the cases when the MS receives signal from the mobile FAP and the macrocellular BS.

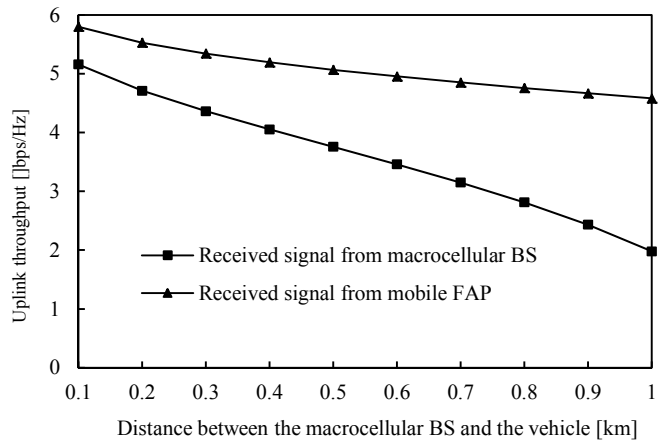

**Fig. 7.** Uplink throughput comparison of macrocellular network between the cases when the MS receives signal from the mobile FAP and the macrocellular BS.

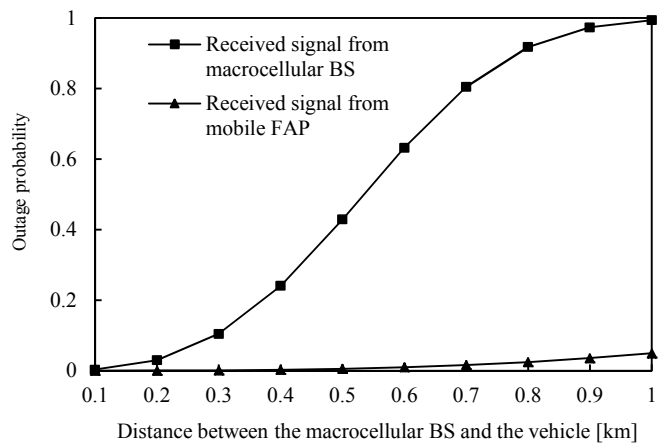

**Fig. 8.** Outage probability comparison between the cases when the MS receives signal from the mobile FAP and the macrocellular BS.

Fig. 7 demonstrates the uplink throughput comparison of macrocellular network. The poor received signal from the macrocellular BS causes very small throughput. However, the received signal level at the receiver outside the vehicle is comparatively stronger. This stronger signal causes the better throughput of macrocellular network. Therefore, the throughput performance can be improved by deploying the femtocellular networks in the vehicular environment.

The MS will be out of connection only if either the link between the macrocellular BS and the outside receiver or the link between the MS and the FAP is disconnected. Fig. 8 illustrates the fact that the outage probability of the proposed mobile femtocellular network deployment is significantly smaller compared to a scheme where MS inside a vehicle receives signal directly from the macrocellular BS. The strong signal between the MS and the FAP causes negligible outage probability. The better *SNIR* level at outside receiver is also improves the outage probability performances. Moreover, the outside receiver is stronger than the MS.

## VI. Conclusions

Femtocells are a novel wireless networking technology that has several advantages including lower cost, better signal quality, and reduced infrastructure cost. The femtocells in the vehicular environment i.e., the mobile femtocells will be the new paradigm of the femtocellular network deployment. The service quality of the macrocellular networks is very poor in the vehicular environment due the long distance between the access network and the MS, the penetration loss, and noise. The successful deployment of mobile femtocellular network will provide the enhanced quality of services for a mobile user inside the vehicle.

In this paper, we discussed the advantages of the mobile femtocellular network deployment. The various service scenarios are presented. We provided a detail analysis of the capacity and outage probability. The simulation results demonstrate that the deployment of mobile femtocells will significantly improve the performance of mobile users in the vehicular environment. Therefore, the femtocellular network deployment in the vehicular networks will be the new paradigm of the femtocell network technology.

## Acknowledgement

This work was supported by Electronics and Telecommunications Research Institute (ETRI).

## References

1. Holger Claussen, Lester T. W. Ho, and Louis G. Samuel, "An Overview of the Femtocell Concept," pp. 221-245, *Bell Labs Technical Journal,* 2008.
2. Mostafa Zaman Chowdhury, Yeong Min Jang, and Zygmunt J. Haas, "Network Evolution and QoS Provisioning for Integrated Femtocell/Macrocell Networks," *International Journal of Wireless & Mobile Networks (IJWMN)*, pp 1-16, August 2010.
3. Vikram Chandrasekhar, Jeffrey G. Andrews, and Alan Gatherer, "Femtocell networks: a survey," *IEEE Communication Magazine*, pp. 59 – 67, September 2008.
4. Mostafa Zaman Chowdhury, Yeong Min Jang, and Zygmunt J. Haas, "Cost-Effective Frequency Planning for Capacity Enhancement of Femtocellular Networks," *Wireless Personal Communications,* DOI: 10.1007/s11277-011-0258-y.
5. http://www.femtoforum.org
6. 3GPP TR R25.820 V8.2.0, "3G Home NodeB Study Item Technical Report," November 2008.
7. Femtoforum, "OFDMA interference study: evaluation methodology document," pp. 9-15, November 2008.
8. Kaveg Pahlavan and Prasant Krishnamurthy, Principles of Wireless Networks, Prentice Hall PTR, New Jersey, 2002.
9. Shaoji Ni, Yong Liang and Sven-Gustav Häggman, "Outage Probability in GSM-GPRS Cellular Systems with and without Frequency Hopping," *Wireless Personal Communication,* pp. 215-234, 2000.